# Enhanced quantum coherence in exchange coupled spins via singlet-triplet transitions

Yujeong Bae,[1,2,3,†] Kai Yang,[2,†] Philip Willke,[1,2,3] Taeyoung Choi,[1,3] Andreas J. Heinrich,[1,3,*] and Christopher P. Lutz[2,*]

[1]Center for Quantum Nanoscience, Institute for Basic Science, Seoul 03760, Republic of Korea.
[2]IBM Almaden Research Center, San Jose, CA 95120, USA.
[3]Department of Physics, Ewha Womans University, Seoul 03760, Republic of Korea.
†These authors contributed equally to this work.
*Corresponding authors: A.J.H. (*heinrich.andreas@qns.science*), C.P.L. (*cplutz@us.ibm.com*)

**Abstract**
Manipulation of spin states at the single-atom scale underlies spin-based quantum information processing and spintronic devices. Such applications require protection of the spin states against quantum decoherence due to interactions with the environment. While a single spin is easily disrupted, a coupled-spin system can resist decoherence by employing a subspace of states that is immune to magnetic field fluctuations. Here, we engineered the magnetic interactions between the electron spins of two spin-1/2 atoms to create a 'clock transition' and thus enhance their spin coherence. To construct and electrically access the desired spin structures, we use atom manipulation combined with electron spin resonance (ESR) in a scanning tunneling microscope (STM). We show that a two-level system composed of a singlet state and a triplet state is insensitive to local and global magnetic field noise, resulting in much longer spin coherence times compared with individual atoms. Moreover, the spin decoherence resulting from the interaction with tunneling electrons is markedly reduced by a homodyne readout of ESR. These results demonstrate that atomically-precise spin structures can be designed and assembled to yield enhanced quantum coherence.

**Introduction**
The coherent control of spin states is a prerequisite for the use of spins in quantum information technologies (*1–3*). However, the quantum properties of spin states in solid-state nanostructures are easily disrupted by interactions with the environment such as electric or magnetic field noise (*4*) as well as unwanted coupling to nearby spins (*5, 6*). To protect the spin states against decoherence, ion traps (*7, 8*), silicon-based qubits (*9*), and quantum dots (*10–12*) adopted particular spin transitions, called 'clock transitions', which are inherently robust against magnetic field fluctuations (*7*). By carefully tuning the parameters in the spin Hamiltonian of a coupled electron-nuclear (*7, 8*) or electron-electron system (*9*), such clock-transition-based spin qubits have been created and shown to be insensitive to magnetic field noise, at least to first order.

To experimentally address sources of decoherence, well-controlled studies of individual spin centers are critical (*13*). Scanning tunneling microscopy (STM) has been intensively used to construct and characterize spin structures (*14, 15*). While the spin relaxation time ($T_1$) of individual atoms (*16, 17*), molecules (*18*), and nanostructures (*19–21*) has been studied using STM, the spin coherence time ($T_2$) of surface atoms is mostly discussed for individual atoms (*17, 22*) and in theoretical works (*23–25*). Recently, electron spin resonance (ESR) in STM has been applied to electrically sense and control individual magnetic atoms on the surface (*26*) as well as interactions between them (*27–29*). Combining the high energy resolution of ESR and the capability of STM to position individual spin centers with atomic precision, ESR-STM now enables the exploration of decoherence in assembled nanostructures.

In this work, we create a two-level system employing magnetic-field-independent spin states of two magnetically coupled spin-1/2 titanium (Ti) atoms. The spacing between the atoms is precisely chosen to create a relatively strong magnetic coupling (~30 GHz) which protects the spin states from fluctuating magnetic fields. The two-level system consists of the singlet and triplet states having magnetic quantum number $m = 0$, and thus it is not sensitive to magnetic field fluctuations to first order (*3*). This gives a spin coherence time that is more than one order of magnitude longer compared to other states in this system of coupled atoms as well as compared to individual Ti atoms. We further improve the coherence time by setting the DC bias voltage to zero to reduce decoherence induced by tunneling electrons (*22*). This is achieved by using homodyne detection, a mechanism previously used in electrical detection of ferromagnetic resonance (*30, 31*) and here applied to ESR-STM.

**Results**
**Spin Resonance of Singlet and Triplet States**
We employed a low-temperature STM that allows imaging, atom manipulation, and single-atom ESR (Fig. 1A) (*26–29*). One or a few Fe atoms were transferred to the tip apex to create a magnetic tip for ESR driving and sensing. We deposited Ti atoms on a bilayer MgO film grown on Ag(001) (see Method section and Supplementary Materials, section S1). On this surface, Ti atoms have two binding sites: on top of the oxygen atom ($Ti_O$) and at the bridge site between two oxygen atoms ($Ti_B$). Both species have a spin of 1/2, most likely due to an attached H atom (*29*) and show negligible magnetic anisotropy, so to good approximation, the spins align to the uniform external magnetic field $\boldsymbol{B}_\text{ext}$ (fig. S2).

We positioned two Ti atoms to form $Ti_O$-$Ti_B$ dimers (Fig. 1A) and characterized the magnetic interactions between Ti atoms using ESR. When two spin-1/2 atoms are magnetically coupled, the eigenstates are given by the singlet ($|S\rangle$) and triplet ($|T_0\rangle$, $|T_-\rangle$, $|T_+\rangle$) states. While two of the triplet states are the Zeeman product states ($|T_-\rangle = |00\rangle$ and $|T_+\rangle = |11\rangle$), the spin-spin interaction causes the superposition of $|01\rangle$ and $|10\rangle$ states and results in the remaining two eigenstates: $|S\rangle = (|01\rangle - |10\rangle)/\sqrt{2}$ and $|T_0\rangle = (|01\rangle + |10\rangle)/\sqrt{2}$. Here, 0 and 1 designate respectively the spin-up and spin-down states of the constituting spins.

Figure 1B shows an ESR spectrum obtained from a $Ti_O$-$Ti_B$ dimer with the atomic separation $r = 0.92$ nm. Four ESR peaks arise from the four transitions that change the total magnetic quantum number $m$ by $\pm 1$ as given in the schematic energy diagram (Fig. 1D). The difference between resonance frequencies ($f_{T_0T_+} - f_{ST_-}$, or equivalently $f_{ST_+} - f_{T_0T_-}$) directly gives the magnetic interaction energy $J$ between the two spins (*29*), which is $0.77 \pm 0.02$ GHz for this dimer spacing (fig. S5). Because the interaction energy $J$ is smaller than the Zeeman energy, the $|T_-\rangle$ state becomes the ground state.

We find that the $Ti_O$-$Ti_B$ dimer can be positioned close enough to yield coupling strengths sufficiently strong to shift the singlet state down in energy to become the ground state (Fig. 1E). Such interaction strength was not accessible for the $Ti_O$-$Ti_O$ dimers (*29*). Decreasing the spacing of the atoms in the $Ti_O$-$Ti_B$ dimer to $r = 0.72$ nm (Fig. 1C) results in three ESR peaks in our measurement range (5–30 GHz). In addition to the two triplet-triplet transitions ($f_{T_0T_+}$ and $f_{T_0T_-}$), the singlet-triplet (S-$T_0$) transition is now visible as resonance at $f_{ST_0} \approx 29$ GHz. Here, the resonance frequency $f_{ST_0}$ directly gives the value of $J$ when the detuning (see below) is negligible.

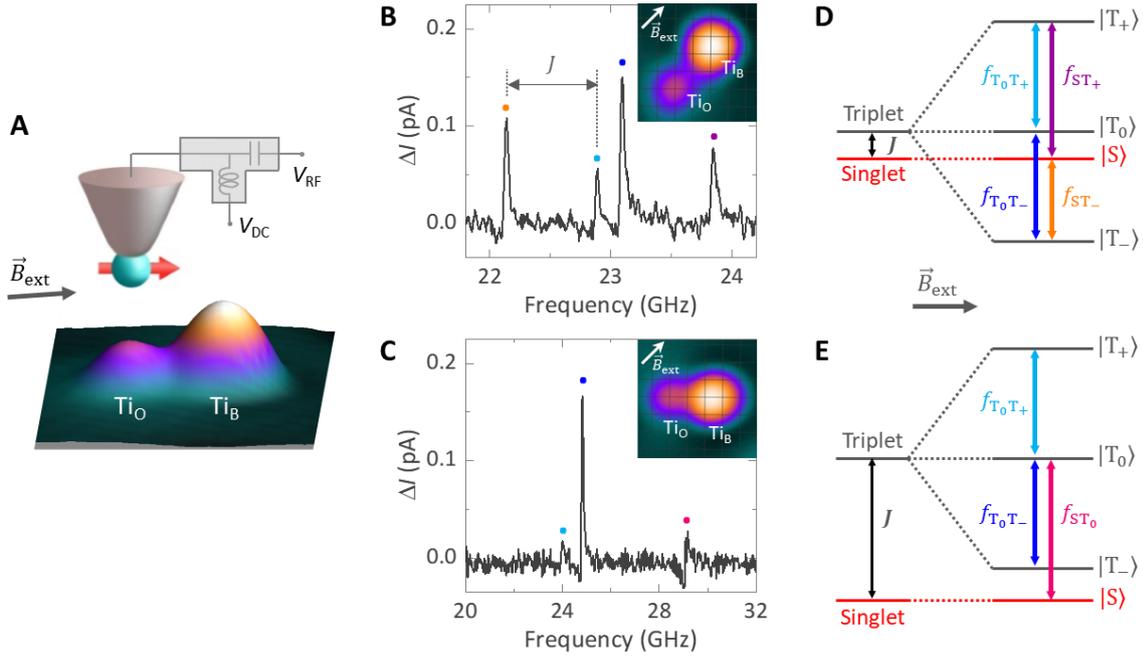

**Fig. 1. Spin resonance for two coupled spin-1/2 Ti atoms.** (**A**) Schematic of the ESR-STM setup with the topographic image of a pair of Ti atoms on a bilayer MgO film, where the Ti atoms are separated by $r = 0.92$ nm. The two species appear with different apparent heights in the STM image: ~1 Å for Ti at the O binding site of MgO (Ti$_O$) and ~1.8 Å for Ti at a bridge site (Ti$_B$) ($V_{DC} = 40$ mV, $I = 10$ pA, $T = 1.2$ K). The external magnetic field is applied almost parallel to the surface. (**B**) ESR spectrum measured on Ti$_O$ in a Ti$_O$-Ti$_B$ dimer with $r = 0.92$ nm and (**C**) $r = 0.72$ nm [$V_{DC} = 40$ mV, $T = 1.2$ K, $B_{ext} = 0.9$ T / (B): $I = 10$ pA, $V_{RF} = 30$ mV / (C): $I = 20$ pA, $V_{RF} = 15$ mV]. Insets: STM images of the Ti$_O$-Ti$_B$ dimer used to measure each ESR spectrum. The grid intersections indicate the positions of oxygen atoms of the MgO lattice. (**D, E**) Schematic energy level diagrams for two coupled spin-1/2 atoms. In (D), the Zeeman energy is larger than the interaction energy $J$ between two atoms, leading to the triplet state as the ground state. In (E), the singlet state becomes the ground state when $J$ is larger than the Zeeman energy. The resonance peaks in (B) and (C) are marked by the same colors as transition labels in (D) and (E), respectively.

### Heisenberg Exchange Coupled Spin-1/2 Ti Atoms

From the ESR peak splitting, we determined the magnetic interaction energy $J$ for 30 dimers with different separations and orientations (fig. S3). The measured values $J$ are given in Fig. 2A as a function of atomic separations ($r$, ranging from 0.72 nm to 1.3 nm). We find that for atomic distances of less than 1 nm, the Ti$_O$-Ti$_B$ dimers are dominantly coupled by the Heisenberg exchange interaction $J\mathbf{S}_1 \cdot \mathbf{S}_2$, where $\mathbf{S}_1$ and $\mathbf{S}_2$ are the spin operators. Moreover, the interaction is found to be isotropic (fig. S3).

The exchange interaction generally shows exponential dependence on the separation between spins (*32*). Given the isotropic interaction energy $J = J_0 \exp[-(r - r_0)/d]$ (*32*) and taking $r_0 = 0.72$ nm, we obtain for Ti$_O$-Ti$_B$ dimers a decay constant $d = 64.6 \pm 4.9$ pm and a prefactor $J_0 = 28.9 \pm 1.3$ GHz. The decay constant matches well with reported values for exchange interactions across a vacuum gap (*29, 33, 34*). For Ti$_O$-Ti$_O$ and Ti$_B$-Ti$_B$ dimers, we obtain $d = 40.0 \pm 2.0$ pm (*29*) and $94.0 \pm 0.3$ pm (fig. S3D), respectively. This difference in decay constant between the dimer types indicates the sensitivity of the exchange interaction to either the orbitals being involved in the interaction or the spatial distribution of spin density (*35*), resulting from the different interaction potentials (*32*) and the different magnetic ground states (*29*). As determined from the intensity of peaks in the ESR spectra (fig. S3) (*27, 29*), $J$ is positive, and thus the coupling between Ti atom spins is antiferromagnetic.

**Energy Detuning of Superposition States**
While a traditional ESR measurement applies a uniform radio-frequency (RF) magnetic field, in the ESR technique used here the RF magnetic field at the Ti atom arises from the modulation of the atom's position (*36*) in the non-uniform magnetic field $\boldsymbol{B}_{\text{tip}}$ (*29, 33*) generated by the STM tip. This tip magnetic field also provides a means to measure the effect of local magnetic fields on the quantum states and their robustness against decoherence. Increasing $\boldsymbol{B}_{\text{tip}}$ by positioning the tip closer to the atom makes the eigenstates deviate from ideal singlet and triplet states. This deviation is related to an energy detuning ε, which is the difference in Zeeman energies between the two atoms. Minimizing this detuning is desired to maximize the quantum coherence of coupled spins, as shown in the following. The detuning arises from two sources: (i) a slight difference in the gyromagnetic ratios $\gamma_1$ and $\gamma_2$ for the two atoms at different binding sites (fig. S2) and (ii) the tip magnetic field which is applied only on one of the atoms (*29*). The Hamiltonian describing the two spins dominantly coupled by the exchange interaction is then given by

$$H = \gamma_1 \hbar\, \boldsymbol{S}_1 \cdot (\boldsymbol{B}_{\text{ext}} + \boldsymbol{B}_{\text{tip}}) + \gamma_2 \hbar\, \boldsymbol{S}_2 \cdot \boldsymbol{B}_{\text{ext}} + J\, \boldsymbol{S}_1 \cdot \boldsymbol{S}_2 \ . \tag{1}$$

Under the approximation that $\boldsymbol{B}_{\text{tip}}$ is parallel to $\boldsymbol{B}_{\text{ext}}$, the energy detuning is given by $\varepsilon = (\gamma_1 - \gamma_2)\hbar B_{\text{ext}} + \gamma_1 \hbar B_{\text{tip}}$ (*29*), resulting in the quantum eigenstates:

$$|T_0(\xi)\rangle = \sin\frac{\xi}{2}|01\rangle + \cos\frac{\xi}{2}|10\rangle$$
$$|S(\xi)\rangle = \cos\frac{\xi}{2}|01\rangle - \sin\frac{\xi}{2}|10\rangle$$

where $\xi$ is a mixing parameter given by $\tan\xi = J/\varepsilon$. When the energy detuning is negligible ($J \gg \varepsilon$), the eigenstates are the ideal singlet and triplet states: $|S\rangle = (|01\rangle - |10\rangle)/\sqrt{2}$ and $|T_0\rangle = (|01\rangle + |10\rangle)/\sqrt{2}$. In contrast, increasing the energy detuning leads to the Zeeman product states: $|01\rangle$ and $|10\rangle$.

The effect of such energy detuning on the eigenstates is a shift of their energy levels, which results in a corresponding ESR frequency shift of the S-$T_0$ transition ($\Delta f_{ST_0}$) from the minimum value of $f_{ST_0}$. Figure 2B shows the measured $\Delta f_{ST_0}$ for Ti$_O$-Ti$_B$ dimers as a function of $\boldsymbol{B}_{\text{tip}}$ for different values of $J$. The minimum in $f_{ST_0}$ is reached where $B_{\text{tip}} = 38 \pm 12$ mT (Fig. 2C). At this field, the detuning is absent, i.e. $\varepsilon = 0$, because the tip field fully compensates the subtle difference in magnetic moments of the Ti$_O$ and Ti$_B$ atoms (fig. S2).

We calculated the eigenvalues and eigenstates (fig. S4) using the Hamiltonian in eq. (1) to fit the experimental results. When $\boldsymbol{B}_{\text{tip}}$ is parallel to $\boldsymbol{B}_{\text{ext}}$, the singlet-triplet energy detuning is given by $\Delta f_{ST_0} = \sqrt{J^2 + \varepsilon^2} - J$. However, it is necessary to account for the tilting of $\boldsymbol{B}_{\text{tip}}$ with respect to $\boldsymbol{B}_{\text{ext}}$ to obtain adequate fits to the ESR frequencies for all allowed transitions in each dimer of Fig. 2. We find that the angles between $\boldsymbol{B}_{\text{tip}}$ and $\boldsymbol{B}_{\text{ext}}$ fall in the range 21–51° depending on the particular tip apex used (Supplementary Materials, section S4).

For weakly coupled dimers ($J < 1$ GHz), a slight increase of energy detuning due to $B_{\text{tip}}$ shifts $f_{ST_0}$ considerably from its minimum value. Using the model Hamiltonian of eq. (1), at the typical tip field of 110 mT, the eigenstates of the dimer with $J = 0.8$ GHz are $|S(\xi)\rangle = 0.92|01\rangle - 0.39|10\rangle$ and $|T_0(\xi)\rangle = 0.39|01\rangle + 0.92|10\rangle$, which more closely resemble the Zeeman product states.

We find that the effects of Zeeman energy detuning ($\varepsilon$) on $\Delta f_{ST_0}$ can be regulated by the coupling strength of two atoms. As shown in Fig. 2C, increasing $J$ to 30 GHz ($J \gg \varepsilon$) markedly reduces the sensitivity of $f_{ST_0}$ on the magnetic field variation. At the same tip field ($B_{tip} = 110$ mT), the eigenstates now remain almost in the ideal singlet and triplet states ($|S(\xi)\rangle = 0.71|01\rangle - 0.70|10\rangle$ and $|T_0(\xi)\rangle = 0.70|01\rangle + 0.71|10\rangle$). Thus, in the following we ensure that $J \gg \varepsilon$ by keeping $B_{tip}$ small ($< 150$ mT), and using large $J$ (30 GHz) and show that this choice results in a decoherence-free subspace.

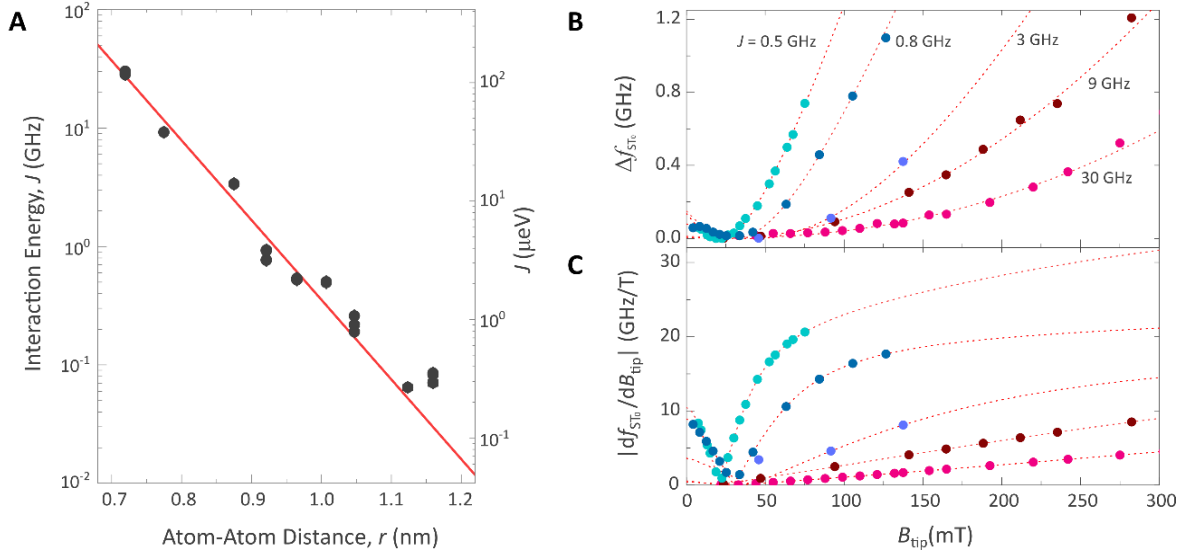

**Fig. 2. Singlet-triplet energy detuning of Ti$_O$-Ti$_B$ dimers with different interaction energies.** (**A**) Magnetic interaction energy determined from ESR measurements for Ti$_O$-Ti$_B$ dimers with different atomic separations. Red lines show the exponential fit, indicative of Heisenberg exchange interaction. The slight deviation of the Ti$_O$-Ti$_B$ interaction energy from the exponential fit is due to the contribution from the dipole-dipole interaction at larger distances. (**B**) The ESR frequency shift of the S-T$_0$ transition ($\Delta f_{ST_0}$) for dimers with different $J$ as a function of the magnitude of tip field, $B_{tip}$. For the dimers with $J = 0.5$, 0.8, and 3 GHz, the resonance frequencies are obtained by $f_{ST_0} = f_{T_0 T_-} - f_{ST_-}$; for the dimers with $J = 9$ and 30 GHz, $f_{ST_0}$ is directly measure from ESR spectra. Strengthening the exchange interaction between Ti atoms protects the $|S\rangle$ and $|T_0\rangle$ states from detuning by $B_{tip}$, reducing $\Delta f_{ST_0}$. (**C**) First-order tip field dependence of $f_{ST_0}$ for the dimers in (B). The clock transitions appear at $B_{tip} = 38 \pm 12$ mT, where $df_{ST_0}/dB_{tip} = 0$.

**Enhanced Spin Coherence Using Magnetic Field Independent States**
Based on the results from the previous sections, we now focus on the spin coherence times of strongly coupled Ti$_O$-Ti$_B$ dimers ($r = 0.72$ nm, $J \approx 30$ GHz). The spin coherence for the singlet-triplet transition and its sensitivity to the external and local magnetic fields are compared in Fig. 3 to (i) the triplet-triplet transition of the same dimer, and (ii) the $|0\rangle$ to $|1\rangle$ transition of an individual Ti$_O$ atom. We obtained the spin coherence time of each transition by fitting the ESR linewidth $\Gamma$ to the Bloch equation model (*26*) (Supplementary Materials, section S5), as a function of RF voltage ($V_{RF}$) (Fig. 3A). In the limit of small $V_{RF}$, the coherence time is given by $1/\pi\Gamma$. This coherence time includes the effect of inhomogeneous line broadening and is designated $T_2^*$ to distinguish it from the intrinsic spin coherence time $T_2$. In our single-spin experiment, inhomogeneous broadening may be due to any time-varying magnetic fields that are present to give temporal ensemble broadening (*37*).

For typical ESR conditions and $B_{tip} = 110$ mT, we find $T_2^* = 99.0 \pm 9.7$ ns for the S-$T_0$ transition (Fig. 3A). Under the same conditions, $T_2^*$ of the triplet-triplet ($T_0$-$T_-$) transition and for the individual Ti$_O$ is ~8 and ~20 times smaller, at only $13.0 \pm 0.3$ ns and $5.3 \pm 0.7$ ns, respectively.

The spin coherence time is closely related to the sensitivity of spin states to the time-varying external and local magnetic fields. The sensitivity to uniform external magnetic fields was characterized by varying the external field magnitude $B_{ext}$ from 0.5 to 1.1 T (Fig. 3B). ESR frequencies $f_{T_0 T_-}$ and $f_{T_0 T_+}$ ($\Delta m = \pm 1$ transition) shift linearly with $B_{ext}$ due to the Zeeman effect on the states $|T_-\rangle$ and $|T_+\rangle$. In contrast, $f_{ST_0}$ ($\Delta m = 0$ transition) shows no Zeeman shift and remains nearly independent of $B_{ext}$, an essential property of a clock transition (7–9).

We now investigate the effect of a local magnetic field by varying $B_{tip}$ over a large range, extending from 10 mT to 0.45 T (Fig. 3C). For the transitions between triplet states, the resonance frequencies $f_{T_0 T_-}$ and $f_{T_0 T_+}$ again increase steadily by the Zeeman energy due to $B_{tip}$ applied to one atom in the dimer. In contrast, $f_{ST_0}$ stays essentially constant when $B_{tip}$ is lower than ~150 mT. The results in Fig. 3 clearly show that the singlet-triplet transition is insensitive to both external and local magnetic field fluctuations, which results in the measured increase of its spin coherence time.

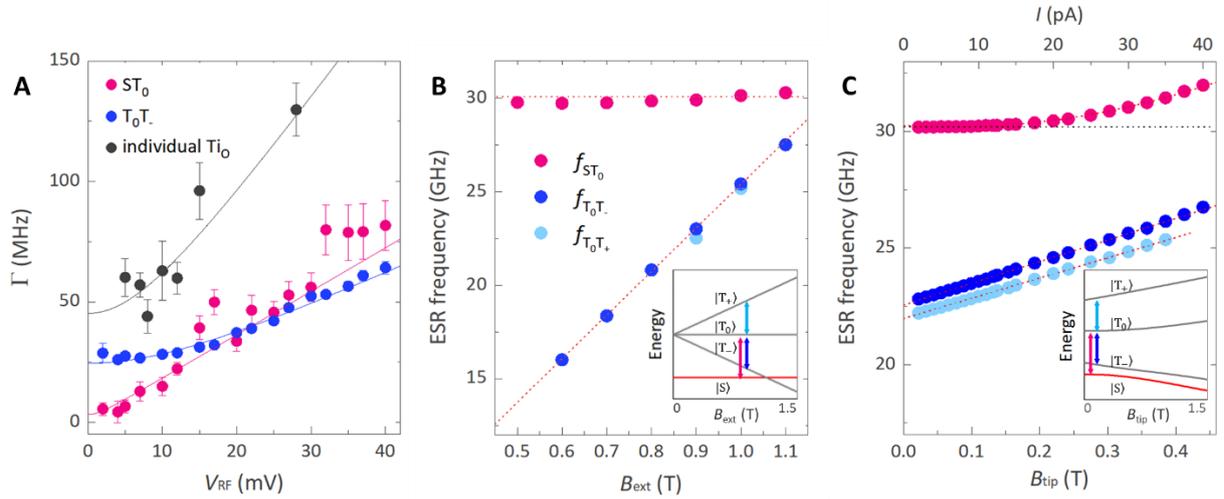

**Fig. 3. Spin coherence of ESR transitions and their sensitivity to external and local magnetic fields.** (**A**) ESR peak width as a function of $V_{RF}$ for the S-$T_0$ and $T_0$-$T_-$ transitions measured on Ti$_O$ in a strongly coupled dimer ($r = 0.72$ nm, $J \approx 30$ GHz), and the $|0\rangle$ to $|1\rangle$ transition of an individual Ti$_O$ atom ($V_{DC} = 40$ mV, $I = 10$ pA, $B_{tip} = 110$ mT, $B_{ext} = 0.9$ T, $T = 1.2$ K). Solid lines are fits to $\Gamma = \sqrt{1 + A V_{RF}^2}/\pi T_2^*$, derived from the Bloch equation model, where the spin coherence time $T_2^*$ is determined by the intercept at $y$ axis and $A$ is a constant. (**B**) ESR frequencies as a function of the external magnetic field $B_{ext}$. For the S-$T_0$ transition, the frequency $f_{ST_0}$ stays almost constant, characteristic of a clock transition. Inset: energy diagram for the four eigenstates at different $B_{ext}$ ($V_{DC} = 40$ mV, $I = 10$ pA, $B_{tip} = 110$ mT, $T = 1.2$ K). (**C**) ESR frequencies as a function of the tip magnetic field $B_{tip}$. $B_{tip}$ is set by the junction impedance ($V_{DC}/I$) and calibrated from the fit (red curves, see also Supplementary Materials, section S4). For the S-$T_0$ transition, the frequency $f_{ST_0}$ stays almost constant and measurably increases when $B_{tip}$ is larger than 150 mT, which reflects the change of eigenstates from the ideal singlet and triplet state. Inset: energy diagram at different $B_{tip}$ ($V_{DC} = 40$ mV, $I = 10$ pA, $B_{ext} = 0.9$ T, $T = 1.2$ K).

## Homodyne Detection as Means to Spin Decoherence Reduction

In addition to magnetic field fluctuations, tunneling electrons are a major source of decoherence of the surface atom's spin in magnetoresistively-sensed ESR (*22*). Here, we show how to achieve further improvements in $T_2^*$ based on the ESR detection mechanism.

In the ESR spectrum of the dimer (Fig. 1C), a notable difference between the singlet-triplet (S-$T_0$) and triplet-triplet ($T_0$-$T_-$) transitions is the lineshape of ESR peaks. For the individual Ti atoms (fig. S2) or triplet-triplet transitions, the ESR signal is nearly symmetric. In contrast, the S-$T_0$ transition appears antisymmetric for low $B_{\text{tip}}$ (Fig. 4A). As the tip field increases, the ESR lineshape becomes more symmetric. Since the ESR detection mechanism depends on the nature of the spin states, such changes in the ESR lineshape for the S-$T_0$ transition in Fig. 4A are a direct consequence of the states changing from the $|S\rangle$ and $|T_0\rangle$ states, towards the Zeeman product states ($|01\rangle$ and $|10\rangle$), as $B_{\text{tip}}$ increases (Supplementary Materials, section S5).

Figure 4B shows the ESR spectra for the $T_0$-$T_-$ and S-$T_0$ transitions at $B_{\text{tip}} = 110$ mT, where the superposition states more closely approximate the $|S\rangle$ and $|T_0\rangle$ states. For the $T_0$-$T_-$ transition, the nearly-symmetric ESR signal results from the change in time-average population of spin states for the atom under the tip (*26*), as detected by $V_{\text{DC}}$. Thus, the peak amplitude decreases with decreasing $V_{\text{DC}}$ (top panel of Fig. 4B). For the S-$T_0$ transition, the time-average population of spin states of the atom under the tip does not vary so it cannot be detected by DC conductance changes. However, the magnetization of the atom along the quantization axis oscillates in time during ESR. The oscillating magnetoconductance at the frequency of the driving voltage $V_{\text{RF}}$ is multiplied by $V_{\text{RF}}$ to produce a DC tunnel current, which can thus be detected. This rectification is known as a homodyne detection (*30, 31*) (for a full description of the ESR lineshape, see Supplementary Materials, section S5). Thus, in the case of the S-$T_0$ transition, both driving and sensing the spin resonance signal can be achieved by employing $V_{\text{RF}}$ only, enabling us to set $V_{\text{DC}}$ to zero. In Fig. 4B, we find that for the S-$T_0$ resonance signal the peak width is narrower for lower $V_{\text{DC}}$. As a result, we find that at $V_{\text{DC}} = 0$, the ESR signal of the S-$T_0$ transition is the sharpest because the tunneling current due to $V_{\text{DC}}$ is absent.

As seen in Fig. 4C, the coherence times $T_2^*$ for all transitions observed increase rapidly with decreasing $V_{\text{DC}}$. Since nearly every tunneling electron induces decoherence of the surface spin (*22*), reducing the number of tunneling electrons improves the spin coherence significantly. At $V_{\text{DC}} = 0$, we obtain $T_2^* = 257 \pm 80$ ns for the S-$T_0$ transition. Note that the ESR measurement at $V_{\text{DC}} = 0$ is only possible for the S-$T_0$ transition (Fig. 4B). Even though we set $V_{\text{DC}}$ to zero, the remaining tunneling current generated by $V_{\text{RF}}$, the finite temperature (*22*), and the relatively short spin relaxation time $T_1$ (*29, 38*) limit the spin coherence time of the S-$T_0$ transition, resulting in the deviation of $T_2^*$ from the reciprocal curve in Fig. 4C.

## Discussion

By controlling the magnetic coupling between electron spins of two atoms, we have demonstrated robust singlet and triplet states and achieved a significantly enhanced spin coherence time. Interestingly, both driving and sensing the singlet-triplet transition do not require a DC voltage, providing an additional way to improve the spin coherence. As a result, we achieved a large improvement of spin coherence by a factor of about 10 compared to the triplet-triplet transition in the same dimer. Moreover, this exceeds the spin coherence time previously determined for individual Fe atoms (*26*), despite the much shorter spin relaxation time $T_1$ for individual Ti atoms (*29*). These engineered atomic-scale magnetic structures may serve as the smallest component for assembling custom spin chains and arrays with enhanced quantum coherence times. The ability of ESR-STM to construct desired multi-spin systems and to electrically access their many-body states might enable the exploration of quantum phases, spintronic information processing, and quantum simulation.

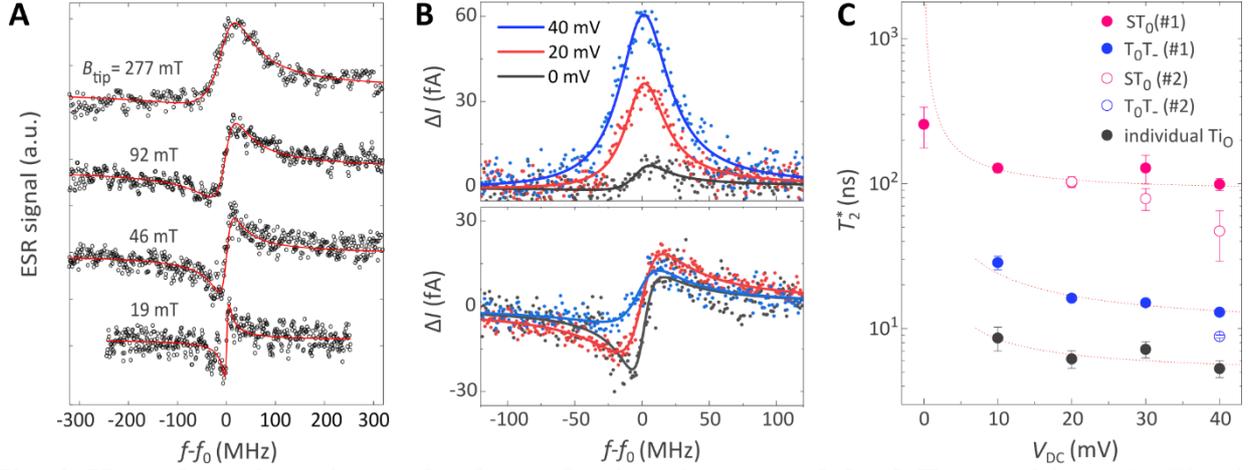

**Fig. 4. Homodyne detection and enhanced spin coherence of the S-T₀ transition.** (**A**) The tip field effects on ESR lineshape of the S-T$_0$ transition. The ESR spectra are normalized and vertically offset. (**B**) DC bias dependence of the ESR signals for the T$_0$-T$_-$ transition (top) and S-T$_0$ transition (bottom). For the S-T$_0$ transition, homodyne detection allows $V_{\text{DC}}$ to be decreased without losing signal intensity ($B_{\text{tip}} = 110$ mT, $V_{\text{RF}} = 20$ mV, $T = 1.2$ K). (**C**) Spin coherence time, $T_2^*$, as a function of $V_{\text{DC}}$. Red curves are reciprocal fit. At fixed junction impedance ($V_{\text{DC}} = 40$ mV, $I = 10$ pA), $T_2^*$ increases with lowering $V_{\text{DC}}$ due to the reduction of tunneling electrons per unit time. For the S-T$_0$ transition, setting $V_{\text{DC}}$ to zero provides further improvement in the spin coherence time by reducing the DC tunneling current. Labels #1 and #2 indicate different dimers (measured with different tips) with same separation ($r = 0.72$ nm) to confirm the reproducibility of $T_2^*$.

## Materials and Methods
### Experimental Design
Experiments were performed using a homebuilt STM system at the IBM Almaden Research Center. We evaporated Ti atoms onto a cold (< 10 K) bilayer MgO film grown on Ag(001). The MgO layer is used to decouple the spin of Ti atoms from the underlying substrate electrons (*39*). Previous works showed that the Ti atoms are likely hydrogenated due to residual hydrogen gas in the vacuum chamber (*29, 40*), and here we denote the hydrogenated Ti atoms simply as Ti. An external magnetic field ($B_{\text{ext}}$) is applied nearly in-plane. An RF voltage $V_{\text{RF}}$ is applied across the tunnel junction for driving spin resonance, and a DC bias voltage $V_{\text{DC}}$ is applied for the DC magnetoresistive sensing of the spin states (Fig. 1A) (*26*).

We thank Bruce Melior for expert technical assistance. We gratefully acknowledge financial support from the Office of Naval Research. Y.B., P.W., T.C., and A.J.H. acknowledge support from the Institute for Basic Science under grant IBS-R027-D1.